\documentclass[doublespacing]{elsart}
\usepackage{graphicx}
\usepackage{amssymb}

\begin{document}
\begin{frontmatter}

\journal{SCES'2001: Version 2}

\title{LiV$_2$O$_4$: evidence for two-stage screening}
\author{ J. Hopkinson}
\author{P. Coleman}
\address{Center for Materials Theory,
Department of Physics and Astronomy, 
Rutgers University, Piscataway, NJ 08854, USA.
}
\thanks{This work was supported by National Science Foundation grant DMR 9983156.  We would like to thank H. Takagi, J. Matsuno, K. LeHur, W. Ratcliff II and C. Hooley for useful discussions.}

%\corauth[1]{Corresponding Author: Rutgers University, 136 Frelinghuysen Rd, P%iscataway, NJ 08854, USA. Phone: (732) 445-4665 Fax: (732) 445-4343}

\begin{abstract}
LiV$_2$O$_4$, a frustrated mixed valent metal ($d^1 \leftrightarrow d^2$), is argued to undergo two spin-screening processes.  The first quenches the effective spin to produce the spin $\frac{1}{2}$ behavior seen below room temperature{\cite{kondo}} , while the second produces the heavy fermi liquid character seen at low temperatures{\cite{urano}}.  We present here a preliminary discussion of a t-J model with strong Hund's coupling of the strongly correlated d-electrons.  Valence fluctuations of the Hubbard operators (S = $\frac{1}{2} \leftrightarrow 1$) combined with the frustration of the underlying corner-shared tetrahedral vanadium lattice are the essential components of our model. 
\end{abstract} 
%%Outline:

\end{frontmatter}

The central question posed to us by LiV$_2$O$_4$ is: how can a mixture of S = 1 and S = $\frac{1}{2}$ quantum spins show behaviour characteristic of localized S = $\frac{1}{2}$ spins over a wide temperature range before condensing to a heavy fermi liquid state?  There have been several attempts to answer this.  Anisimov et al{\cite{anisimov}} have postulated the existence of two species of electrons and invoked the exhaustion phenomenon to explain the low temperature fermi liquid phase.  Varma{\cite{varma}} proposed that the spin $\frac{1}{2}$ state arises due to mixed valence but it is not immediately clear how this can give rise to a second stage spin quenching process.  Kusunose et al{\cite{kusunose}} have considered a two-band Hubbard model in the limit of small Hund's coupling.  Fulde et al {\cite{fulde}} have considered each magnetic tetrahedron to be occupied by two S = 1 and two S = $\frac{1}{2}$ spins, the latter forming gapless Heisenberg chains similar to a Luttinger liquid at low temperatures.  And recently, Burdin et al {\cite{burdin}} have proposed a single-band Kondo lattice model on a frustrated lattice.  We present here a provisional description of a simple two-band model which we believe captures the essential physics behind these two stages (Fig.1).

%our model--the physics behind it (the gist).

The essence of our model is a high-temperature mixed valent spin-quenching which allows a symmetric spin channel to become heavy free electrons.  At low temperatures, these free electrons strongly couple in an orthogonal channel{\cite{coleman&tsvelik}} via a Kondo-like term that arises from antiferromagnetic interactions between the vanadium atoms--which is not permitted to win energetically due to the underlying frustration of the lattice, hence a very heavy fermi liquid of holes forms at low temperatures.  

%our model--explained
We consider a hole-doped antiferromagnetic insulator on a frustrated lattice which we model as a spin 1 t-J model with strong Hund's coupling.
The crystal fields present in the trigonally distorted cubic environment of the vanadium atoms mean that in the real system there are three orbitals of interest.
These can be classified as of degenerate $e_{g^{'}_{\pm}} = \sqrt{\frac{5}{4 \pi}}(\pm(\frac{x^2 - y^2}{2} - \frac{xz}{\sqrt{2}}) + i(xy + \frac{yz}{\sqrt{2}}))$ and $a_{1g} = \sqrt{\frac{5}{8 \pi}}(3 z^2 - r^2)$ symmetries where the z axis has been taken to lie along [1,1,1] and the y axis along [1,-1,0].  It is a subtle matter{\cite{matsuno,ueda}} which of these lies lower in energy, and for simplicity we will restrict our discussion to a hole doped band of $e_{g^{'}_{+}}$ symmetry and lower energy $a_{1g}$ in agreement with most of the community.{\cite{anisimov,kusunose,singh}} 

The Hamiltonian corresponding to this picture is:
\begin{equation}
H =  -\sum_{<ij>}t_{ij}X_{0 \sigma}^iX_{\sigma 0}^j +  J_{Heis}\sum_{<ij>} {\bf{s}}_i \cdot {\bf{s}}_j
\end{equation}
where $X_{0 \sigma} = b^{\dagger}d_{1 \sigma}$ is a Hubbard operator allowing the holes to move, $t_{ij} \approx \psi_{eg_{+}^{'}}^{*}(x-x_i)\psi_{eg_{+}^{'}}(x-x_j)$ is the hopping amplitude, ${\bf{s}} = \pmatrix{d_{1 \alpha}^{\dagger} & d_{2 \alpha}^{\dagger}}\frac{\sigma_{\alpha \beta}}{2}\pmatrix{d_{1 \beta} \cr d_{2 \beta}}$ is the total spin, and we need to enforce the constraints $n_{d_1} + n_b = 1$, $n_{d_2} = 1$, (U $\rightarrow \infty$)  and $S^2 = 2$ (J$_{Hund's} \rightarrow \infty$).  

Below T$_1 \approx t|b|^2$, $\langle b \rangle \ne 0$ and the holes begin to couple strongly to the lattice, freeing spins as they do, until eventually all $d_1^{'}$ spins are free to move leaving behind S = $\frac{1}{2}$ at each site.  Physically we expect to be left with a quarter-filled fermi sea of mobile electrons.  Then at high temperatures we expect a wavefunction of the form 
\begin{equation}
|\psi> = {\mathcal P} \prod_{k < k_F} d_{ k \sigma}^{\dagger (1)} \prod_j d_{ j \sigma}^{\dagger}|0>
\end{equation}
where ${\mathcal P} = \prod_{i \alpha} P_{H}P_{G_{\alpha}}$ is a projection operator where $P_H = (\frac{3 + b^{\dagger}b}{4} + {\bf{s}}_1(i)\cdot{\bf{s}}_2(i))$ projects out singlet states, and $P_{G_{\alpha}} = (1 - n_{\alpha \uparrow}(i)n_{\alpha \downarrow}(i))$ is a Gutzwiller projection.
%Discussion of Fourier-transforming, why local & itinerant....

Using a Hubbard-Stratonovich transformation, we can decouple the quadratic
terms of the Heisenberg interaction to write a mean field Hamiltonian:

\begin{eqnarray}
H &=& \sum_k \epsilon_k d_{k \sigma}^{(1) \dagger}d_{k \sigma}^{(1)} + \pmatrix{d_{1 \sigma}^{\dagger} & d_{2 \sigma}^{\dagger}}\pmatrix{\lambda_1 & i \xi \cr i \bar \xi & \lambda_2} \pmatrix{d_{1 \sigma} \nonumber \cr d_{2 \sigma}} + \lambda_1(b^{*}b -1) - \lambda_2 \\ &+& \sum_k(\bar \Delta \phi^{* (2)}_k d^{\dagger}_{1 k \beta} d_{2 \beta} + h.c.) + 2 \frac{\bar \Delta \Delta}{J_{Heis}}
\end{eqnarray}
where Lagrange multipliers $\lambda_1$, $\lambda_2$ and $\xi$ enforce $U \rightarrow \infty$ and $J_{Hund's} \rightarrow \infty$ respectively, $\epsilon_k = -t_k b^*b \phi_k^{(1)}$ and $\Delta$ is an RVB-like gauge field{\cite{anderson}} that we fix by taking a constrained saddle-point approximation.  At high temperatures, $\Delta = 0$ is a stable saddle-point and the transition temperature below which the Kondo effect becomes apparent is governed by the condition
\begin{equation}
\frac{\partial^2{\mathcal{F}}}{\partial \bar \Delta \partial \Delta}|_{<d_{1 \alpha}d_{2 \alpha}^{\dagger}> = 0} = F_{\bar \Delta \Delta} + \frac{2}{J_{Heis}} - \frac{F_{\bar \Delta \xi}F_{\bar \xi \Delta}}{F_{\xi \xi}} =  0
\end{equation}
where $\frac{\partial {\mathcal{F}}}{\partial \Delta}\delta \Delta + \frac{\partial {\mathcal{F}}}{\partial \xi}\delta \xi = 0$ has been used to express results in terms of unconstrained second derivatives and we have written the full free energy as ${\mathcal{F}} = F + \frac{2 \bar \Delta \Delta}{J_{Heis}}$ to explicitly show that $\frac{1}{J_{Heis}^{*}} = \frac{1}{J_{Heis}} - \frac{F_{\bar \Delta \xi}F_{\bar \xi \Delta}}{2 F_{\xi \xi}}$.   Since $F_{\bar \xi \xi}$ is in general a negative function, overlap functions $\phi^{* (2)}_k$  for which $F_{\bar \Delta \xi} \ne 0$ experience a downward renormalization of their effective Kondo interaction, while channels{\cite{coleman&tsvelik}} with $\sum_k \phi^{* (2)}_k = 0$ will remain strongly interacting as $J_{Heis}$ becomes the dominant scale in the problem.  That one cannot form two spin singlets in the same orbital channel is a consequence of the Pauli exclusion principle.
Then, a second stage spin-quenching occurs in this anti-symmetric channel leaving us with a heavy fermi liquid ground state of holes with wavefunction
\begin{equation}
|\psi^{'}> = {\mathcal P}\prod_{k<k_F}(d_{k \sigma}^{\dagger (1)} + \beta d_{k \sigma}^{\dagger (2)})|0>
\end{equation}
where $\beta$ has the symmetry of the second channel.

%conclusions:symmetry of overlap-deHaas van Alphen, R_H sign change, two-stage
%first mixed valent ~ 550 K, second T_K ~ 50K quenches spin by 4K
We have proposed a toy t-J model of LiV$_2$O$_4$ which includes a mixed valent spin-quenching at high temperatures ($\approx 400 K$ experimentally{\cite{muhtar}}), which precedes a low temperature Kondo-like screening with $T_K^{*} \approx t|b|^2e^{-\frac{t|b|^2}{J_{Heis}}} \approx 50K$ leaving a heavy fermi liquid state by T = 4K.  That this state is energetically realizable at all is due to the frustration of the lattice. (The spin-1 Mott insulator, ZnV$_2$O$_4$ undergoes a phase transition at $\approx 40K$ to an antiferromagnetic ground state{\cite{ueda}}.)  The symmetry of the overlap functions $\phi_i^{(1)} = \frac{1}{\sqrt{6}}(1,1,1,1,1,1)$ and $\phi_i^{(2)} = \frac{1}{\sqrt{3}}(1,\frac{-1}{2},\frac{-1}{2},1,\frac{-1}{2},\frac{-1}{2})$  may be measureable by de Haas-van Alphen experiments, and the sign of the Hall constant {\cite{urano}} is consistent.  Detailed results will follow in a future paper{\cite{hopkinson}}.

Since writing we have learned of a recent two band model by Lacroix{\cite{lacroix}}, with a superexchange coupling localized electrons and a finite Hund's exchange coupling to itinerant electrons.

%\end{fmffile}
\begin{figure}
\includegraphics[scale=1.3, angle = 90]{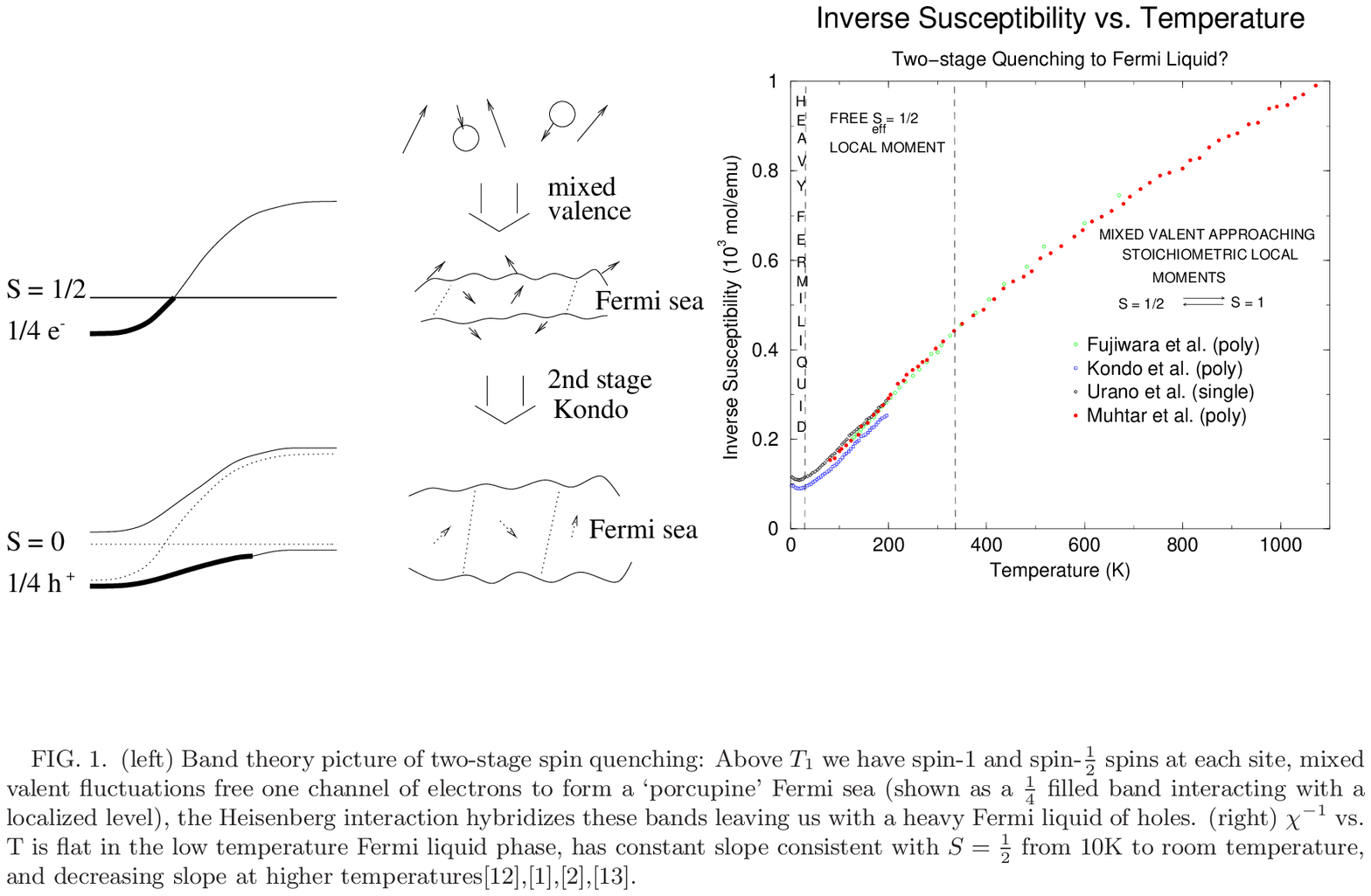}
%\includegraphics[scale=0.8]{bandfs.eps}
%\caption[angle = 90]{\label{figure}(top) Band theory picture of two-stage spi%n quenching: Above $T_1$ we have spin-1 and spin-$\frac{1}{2}$ spins at each %site, mixed valent fluctuations free one channel of electrons to form a `porc%upine' fermi sea (shown as a $\frac{1}{4}$ filled band interacting with a loc%alized level), the Heisenberg interaction hybridizes these bands leaving us w%ith a heavy fermi liquid of holes. (bottom) $\chi^{-1}$ vs. T is flat in the %low temperature fermi liquid phase, has constant slope
% consistent with $S = \frac{1}{2}$ from 10K to room temperature, and decreasi%ng slope at higher temperatures.[12],[1],[2],[13]}
%{\cite{fujiwara,kondo,urano,muhtar}}}
\end{figure}

\end{document}